# The knowledge acquisition process from a complex system perspective: observations and models

**Fátima Velásquez-Rojas,** *Instituto de Física de Líquidos y Sistemas Biológicos (UNLP-CONICET) & Departamento de Ciencias Básicas, Facultad de Ingeniería, Universidad Nacional de La Plata (UNLP), B1900BTE La Plata, Argentina.*

**María Fabiana Laguna,** *Centro Atómico Bariloche, Consejo Nacional de Investigaciones Científicas y Técnicas (CONICET) & Profesorado en Física, Universidad Nacional de Río Negro (UNRN), R8402AGP Bariloche, Argentina.*

RUNNING HEAD: Knowledge acquisition as a complex system

Correspondence should be addressed to: Fátima Velásquez-Rojas, Instituto de Física de Líquidos y Sistemas Biológicos (UNLP-CONICET), B1900BTE La Plata, Argentina. E-mail: fatima.velasquez@ing.unlp.edu.ar

*Abstract:* In this work we study the knowledge acquisition process in a teaching-learning scenario that takes place within the classroom. We explore two complementary approaches, which include classroom observations and student surveys, and the formulation of theoretical models through the use of Statistical Physics tools. We develop an analytical model and a set of dynamics agent-based models that allow us to understand global behaviors, as well as to follow individual trajectories in the knowledge acquisition process. As a proxy of the final achievements of the students we use their final grade, allowing us to assess the validity of our approach. Our models, supported by observations and surveys, reproduce fairly well the process of acquiring knowledge of the students. This work sheds light on the internal dynamics of the classroom and allows us to understand some global aspects of the teaching-learning process.

**INTRODUCTION**

Human beings are constantly involved in social relationships in the educational context. These relationships generate a wide range of emotions and experiences that contribute directly or indirectly to the acquisition of knowledge, concept addressed, among others, by the philosopher J. Locke (Locke, 2001). Some authors proposed that we acquire knowledge through experiences (Mack & Meadowcroft, 2009) and hence the process of acquiring knowledge should be defined as learning through experiences and experiments (Mathew, 1985). Researchers and academics have used a variety of tools to research and measure knowledge acquisition and knowledge creation in different study populations, such as teachers, students, and faculty members (Kaba & Ramaiah, 2019). Different groups were found to acquire knowledge for different purposes through different sources and channels. The tools designed include questionnaires, interviews, observations and content analysis, among others. However, the strategy of integrating data with mathematical models is, in general, much less explored and is the aim of this work.

In a classroom, a series of individual and social processes take place that are integrated into another of greater importance: education. The emerging dynamics undoubtedly indicates that we are facing a system made up of individuals who influence each other, and the result of such interaction cannot be explained as the sum of the isolated behavior of individuals. Thus, when talking about systems in the context of human behavior, as occurs within a classroom, the link with Complex Systems naturally emerges (Guastello & Gregson, 2011; Lemke & Sabelli, 2008; Ricca, 2012; Quezada & Canessa, 2008). This new understanding of phenomena in nature is a promising field of multidisciplinary research that has attracted the attention of a growing number of scientists, especially in recent decades.

The exchange of ideas in the classrooms varies according to the actors of the system and, therefore, the thematic and cultural context in which the relationships are woven. The joint construction of meanings through interactions in the classroom has been studied mostly from the sociocultural cognitive approach, according to which learning and the acquisition of knowledge result from social interaction. This approach is imposed as one of the most robust contemporary paradigms when it comes to characterizing the dynamics of the learning interaction between teachers and students and between students (Mercer and Howe, 2012).

The teaching-learning processes are quite complex, and their study has become a very active area of research (Berliner & Calfee, 1996; Lemke & Sabelli, 2008; Stamovlasis & Koopmans, 2014). The first studies carried out by psychologists and sociologists, express that such processes involve all kinds of social activities (Piaget, 1929; Vygotsky, 1978). Several mechanisms that study human behavior have been proposed in the social science literature (Latané, 1981; Latané & Nowak, 1997). In fact, these mechanisms go hand in hand with cognitive psychology which is dedicated to the study of human behavior that focuses on the unobservable mental aspects that mediate between the stimulus and the open response and which has been used in the field of Education as one part of a solution that involves helping students to better regulate their learning through the use of effective learning techniques (Dunlosky, 2013). Cognitive psychology uses methods such as systematic observation, experimentation and even computer simulation of cognitive processes, distinguishing their constituent elements and their dynamics over time. Since learning, as a cognitive process, has incredibly complex behavior, it must cover topics such as exploring the interface between nonlinear dynamic systems (NDS). That is why the NDS offers a new perspective on educational reality (Stamovlasis & Koopmans, 2014). Even when in the social sciences area talks about educational complexity, a sufficiently robust theoretical-methodological framework has not been established to characterize the dynamics of class interactions in the classroom as a complex system (Di Paolo & De Jaegher, 2012). However, the complexity of educational acts goes from local dynamics typical of the human beings that compose them to global dynamics that integrate the interaction of the agents that are part of the process.

An educational setting is a global dynamic network that is sustained by local interactions that are modulating the teaching-learning process; however, the evolution of the system is subject to global and multivariate changes that are not possible to establish as univocal relationships restricted to the specific action of its components (Scott, 2008). The link between Education and Dynamical Complex Systems is addressed very clearly by Koopmans & Stamovlasis (2016), from which we reproduce some of the main ideas. The word "complex" should be understood as that the behavior of a larger systemic constellation cannot be readily reduced to that of its individual members, in other words, the whole is greater than the sum of its parts. This means that we can understand student learning in terms of collaborative behavior in the classroom in which it takes place, while a classroom climate conducive to learning cannot be readily reduced to the learning or interactive behavior of individual teachers and students. "Dynamical" refers to that current behavior is understood in terms of deviations from past behavior. As a result, the perspective focuses on behavioral change and its determinants, rather than on outcomes frozen in time. The authors finally emphasize that we should be concerned with individual learning trajectories rather than whether students achieve certain benchmarks or performance goals as a group. This connects us with the approach we choose to address this issue.

The possibility of approaching education from the perspective of complex systems is perhaps the main cause of the growing interest in this topic by the community of statistical physicists, due to the fact that some of the proposed teaching-learning mechanisms can be placed within the framework of the dynamics of interacting particles, and can be modeled using the methods and tools developed during the last decades. These methods of analysis of modern physics allow deducing emergent properties from the microscopic interactions of the particle system, which is very useful when studying larger systems. The theoretical and computational models that are commonly used belong to the family of agent-based models (Davidsson, 2002), in which the social behavior of people is, in fact, modeled as a system of interacting particles.

In agent-based modeling (ABM), a system is modeled as a collection of autonomous decision-making entities called agents. Each agent individually assesses its situation and makes decisions on the basis of a set of rules (Bonabeau, 2002). The benefits of ABM over other modeling techniques can be captured in three statements: (i) ABM captures emergent phenomena; (ii) ABM provides a natural description of a system; and (iii) ABM is flexible. It is clear, however, that the ability of ABM to deal with emergent phenomena is what drives the other benefits. There are some issues related to the application of ABM to the social sciences as education because they involve behaviors and decisions difficult to quantify, calibrate, and sometimes justify. However, ABM is the only option to deal with such situations computationally. In our case and since the interactions between the agents are complex and nonlinear, and the population is heterogeneous, with individuals (potentially) different, such models become an essential tool when describing the observed behaviors, but more importantly, to predict results in alternative scenarios.

The starting point of this work are the models proposed by Bordogna and Albano (2001, 2003), who showed that the structure of collaborative groups formed by students can influence their achievements. Our interest in these studies stems from the fact that the processes of learning and understanding of physics and mathematics are based on well-defined conceptual frameworks.

In this sense, implicit norms have been identified in the classrooms: regulations, conventions, morals, truths, and instructions that have been studied in particular environments such as mathematics, whose communication is very specific (Cobb et al., 1992). In another work, Paul Cobb, Erna Yackel and Terry Wood (1992) analyzed the importance of the dynamics that takes place in the classroom in a Mathematics course. They state that the learning-teaching process is constituted interactively in the classrooms through individual and collective activities between teachers and students. This suggests a cognitive and sociological analysis that represents a complex conceptual framework of life in the classroom.

In this work, we study the knowledge acquisition processes in a teaching-learning scenario that takes place inside the classroom using two different and complementary approaches: one includes classroom observations and surveys; the other is the dynamic modeling with tools of Statistical Physics, through the development of agent-based models and stochastic numerical simulations. In this framework, our contribution to this research area is to analyze the educational process as a dynamic complex system to better understand the process of knowledge acquisition, coupling quantitative and qualitative strategies that integrate knowledge acquisition measures with numerical simulations. The proposed approach focuses on the internal processes in the classroom and allows us to understand some global aspects of the teaching-learning process.

The article is organized as follows: the Method Section is divided into 2; in Method 1 are described the participants and its educational context, the measures (which include the materials used to construct our Knowledge Acquisition model) and the procedure where it is explained in detail the surveys and educational observations we use to build such a model. In Method 2, we introduce the Agent-Based model formalism and describe different versions for the acquisition of knowledge, as a more general and complementary framework for this kind of processes. Then, we present the main results of this work and finally, we summarize and discuss our findings.

# METHOD

## Method 1: Knowledge acquisition model

**Participants**

This research was carried out with several sections of students who attended the Physics II course in the Faculty of Engineering of the National University of La Plata (UNLP), Argentina.

The Physics II course is taught in the second year of the career, specifically in the third semester of all Engineering careers. It is important to comment that the Faculty offers 13 engineering degrees, so the interest of the students in the course can vary greatly. In Physics II, concepts seen in these previous courses are recovered and used, providing new essential physical concepts for the training of the future engineer.

The classes are theoretical-practical and are held twice a week in sessions of 4 hours per class. The course is divided into two parts, each part having a written partial test. The score of these evaluations can range from 0 to 10. The course approval regime is by "direct promotion", which implies being exempt from the final test (if the average between the two partial exams is 6 or more) or promotion by final test (if the average is between 4 and 6). The partial tests have two instances of recuperation: a makeup exam within the semester, and a "floating" exam at the end of the semester, where the student can improve any of the lowest scores obtained in previous tests.

The research was done during two semesters in four different sections, where we were able to carry out the classroom observations, surveys, and follow the learning process of 81 students, from whom we have the final grade they obtained in the course.

Following the ideas proposed by Bordogna and Albano (2001) we classified the group of students into three different sets, according to their final achievements $K_f$, that we relate to the final grade obtained in the course.

The group classification of students was done according its $K_f$ as follows: (a) High-achieving (HA) students: $8 \leq K_f \leq 10$, (b) Average-achieving (AA) students: $6 < K_f < 8$ and (c) Low-achieving (LA) students: $K_f \leq 6$.

In Fig. 1 we show the final grades, which from now on we will call *Real $K_f$*, obtained for the *81* students that we include in the present work. These data provide us with the necessary information to contrast our theoretical models.

<<Insert Figure 1 About Here>>

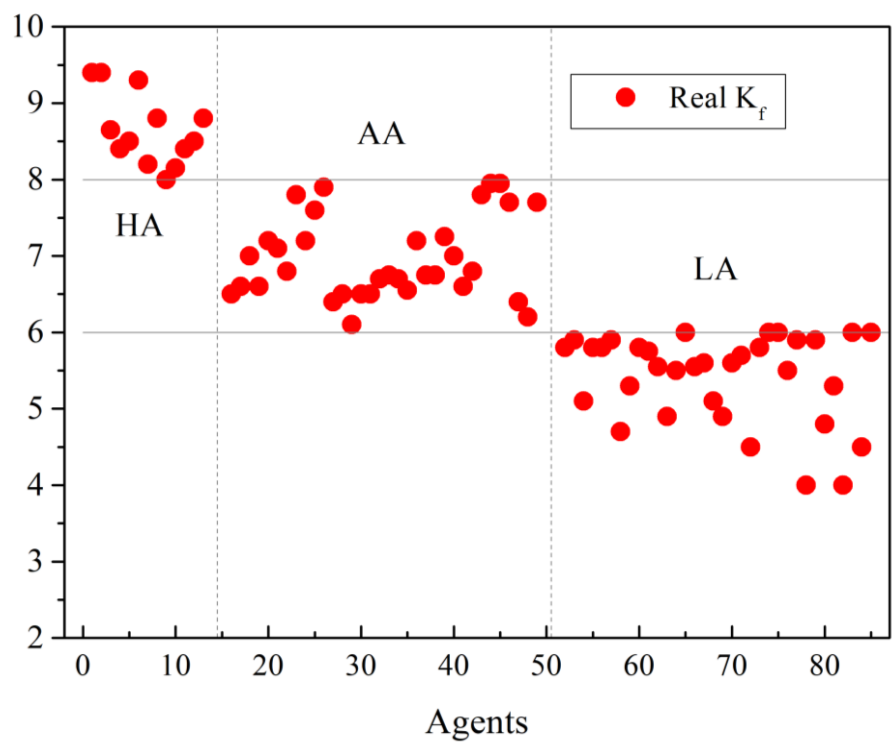

Fig. 1. Group classification of students into three different sets according to their final grade as High-achieving (HA) students, Average-achieving (AA) students and Low-achieving (LA) students for *N=81*. The vertical lines separate the three datasets, while the horizontal lines indicate the scores that we used for the classification made. The data shown are sorted randomly for each group.

Interestingly, we found that each of these groups has qualitatively different characteristics regarding the relevance of the factors considered in the construction of the new knowledge, as we will discuss further. Using the features observed in each group of students during the realization of this work, we propose an analytical model to explain the acquisition of knowledge. We would like to emphasize that in our study we focus on a specific type of learning, related to scientific concepts of classical physics. While we are aware that this is not the only value learned in the classroom, we simplify the concept of knowledge to use the final grade as a concrete and quantifiable measure of student performance.

**Measures**

As we said, the knowledge acquisition process is extremely complex and depends on many variables. In the context of complex systems, the acquisition of new knowledge on a given course by a given individual can be thought as a dynamic variable influenced by several extrinsic and intrinsic factors (Bordogna & Albano, 2001). The first ones are those related to the interactions that the individual has with their environment (teachers, peers), while intrinsic factors can be associated with aptitude (understood as an ability to do something), personal motivation, and prior knowledge that the individual possesses.

We collected information among students about the previously mentioned factors, through surveys (of own authorship) in which we analyze their levels of: aptitude ($A$) with questions to know abilities or performances in past or present courses with objective character; personal motivation ($M$) with questions to know experiences and expectations in relation to the course; prior knowledge ($K_0$), influence of the teachers ($T$) and influence of peers ($P$). We also include information on the spatial distribution of the students and the formation of groups ($G$), obtained through direct observations of the classes. These data were integrated and the assigned numerical value was compared with the final grade obtained in the course (as we said, the score of these evaluations can range from 0 to 10).

In order to explicitly compare the classroom observations and surveys with the final achievement of the students, we write the final knowledge on a given topic as the sum of a set of contributing factors. For a student $i = 1,2,3, ... N$, its final knowledge $K^i_f$ can be posed as:

$$K^i_f = \left(\sum_{j=1}^{3} \alpha_j M_j\right)^i + (A + K_0)^i + (G + P)^i + \theta^i T^i \tag{1}$$

where α is a coefficient related to the self-perception of the student performance over the course of the semester and modulates the motivation (*M*), and θ weights the term related to teacher interaction (*T*) and expresses the perception of the benefit received from the different strategies used in this regard. Both coefficients (α and θ) are normalized to *1* and were obtained through specific questions in the surveys presented in Table 1.

Besides, the contribution of the pairs to the acquisition of knowledge is composed of two factors: the group conformation *G* and the pair interaction *P*. Note that *G* is the only factor that arises not from the survey but from the observations made during the semester, and gives us information about the student´s strategies used to carry out peer interaction. *G* and *P* together measure the relevance of the interaction between pairs.

We would like to emphasize that, although the equation in the KA model is linear, the term of "Pairs" can be interpreted as an effective version of a real non-linear interaction. As will become clearer when we come to the description of the Agent-Based models, this term is actually non-linear, and its detailed description can be made with this type of approach. However, the simplification made in the KA model remains valid in light of the results obtained.

**Surveys.** The following are the surveys carried out on students during the semester of classes. The numbers or letters between brackets correspond to the value that we assign to each of them, in order to transfer the answers to the KA model of Eq. 1.

---

<<Insert Table 1 About Here >>

Table 1.

Surveys carried out on students during the semester of classes.

| Survey | Quantities | Item | Options | Values |
|---|---|---|---|---|
| 1 (first day of the course) | $A$ | 1) At the end of the Physics I course: | I was excepted from the final test<br>I went to the final test<br>I failed | 2<br>1<br>0 |
| | $M$ | 2) At the beginning of Physics II, what is your level of expectation?<br><br>Because the course: | Much<br>Intermediate<br>None<br><br>Excites me<br>It is a requirement | 1<br>0,5<br>0<br><br>1<br>0 |
| | $K_0$ | 3) When are the electric and magnetic fields related?<br><br>Does the electrical capacity just depend on the geometry of the object? | Always<br>Sometimes<br>Never<br><br>Yes<br>No | 0<br>1<br>0<br><br>1<br>0 |
| 2 (end of the first part of the course) | $M$ | 1) So far, describe your experience in Physics II: | I really like it<br>I like it<br>It is indifferent to me<br>I do not like it | 1<br>0,5<br>0<br>-0,5 |
| | $\alpha$ | 2) At the end of Module 1 and in level of difficulty, Physics II has been: | Very difficult<br>Difficult<br>Intermediate<br>Easy | 0,25<br>0,50<br>0,75<br>1,00 |
| | $K_0$ | 3) Are the Kirchhoff Laws valid in alternate current circuits?<br><br>The image of a convex mirror is always: | Yes<br>No<br><br>Real<br>Virtual | 1<br>0<br><br>0<br>1 |
| 3 (end of the course) | $M$ | 1) At a general level, describe your experience in Physics II: | I really liked it<br>I liked it<br>I was indifferent to me<br>I did not like it | 1<br>0,5<br>0<br><br>-0,5 |
| | $T$ | 2) Were the lectures useful for you? | Yes<br>Little | 1<br>0,5 |
| | | Was the interaction with the rest of the teaching team useful to you? | No | 0 |
| | | | Yes | 1 |
| | | | Little | 0,5 |
| | | | No | 0 |
| | *P* | 3) At a general level, describe your way of studying: | Alone | 0 |
| | | | In group | 0,5 |
| | θ | 4) At the time of study, which activity was the most beneficial for you? (You can check several options) | Lectures | A |
| | | | Consulting hours with my group | B |
| | | | Office hours | C |
| | | | Private tutoring | D |
| | α | 5) Do you think you have acquired the minimum knowledge required by this course? | No | 0,33 |
| | | | Little | 0,67 |
| | | | Yes | 1,00 |

--------------------------------------------------------------------------------

The combination of strategies for the question that measures θ was given the following numerical values: ABC, AB, AC, BC=1; A, B=0,7; C, AD, BD, ABD, ACD, BCD = 0,5; CD=0,3; D=0,1 (they could mark several options). These values were given to encourage the use of the strategies provided by the specific section to which the students belonged (options A, B). It is worth remembering that theta represents the strategy or strategies to carry out the interaction with the teacher.

In the next subsection we describe in more detail the way in which surveys and observations were incorporated to the model described by Eq. 1 and propose an agent-based model for the acquisition of knowledge as a complementary framework for this work.

**Procedure**

The information collected to test the validity of Eq. 1 was carried out in two different ways. One corresponds to classroom observations such as the number of students that started and finished the process at each course, age range, gender and spatial distribution. It was particularly helpful to register the location of the students in the classroom to detect the formation of clusters (understood as a group of people who sit and work together).

Besides, we carried out the surveys presented in the previous section, that were used to determine the different quantities involved in our model. The three surveys were made in different moments of the semester: the first, at the beginning of the semester, the second at the end of the first part of the course, and the last one after the end of the semester. The surveys contained closed-ended questions to quantify each term of Eq. 1: self-perception in relation to the course, expectations, preferences, feelings during the process, prior knowledge and skills, interaction with peers and teachers and the strategies that they used to carry out such interaction. The numerical values assigned to the responses were presented in Table 1, and this information was compared with the final score of each individual.

It is worth noting that the values assigned to the different quantities are arbitrary. The Aptitude varies between $0 \leq A \leq 2$, the Motivation is in the range $-1 \leq M \leq 4$, and Peers and Teachers can vary between *0* and *2*. The terms and coefficients that are asked more than once in the surveys ($\alpha$, $M$ and $K_0$) are added up to obtain the $K^i_f$. Besides, although motivation $M$ was asked in the 3 surveys, for the first survey we made its coefficient $\alpha_1 = 1$, since it represents a self-perception of the student performance in the course and at that time they were starting.

As already mentioned, each of these groups has different characteristics regarding the relevance of the factors considered in the construction of $K^i_f$. Moreover, the results of the model must be compared with the final scores that range between 0 and 10. For these reasons we propose a different fit for each group according to their observed characteristics, and we choose the values that best fit the actual data. Such setting values are the coefficients $f^X$ that appear before each term of the following expression:

$$K^i_f = f^X_M \left(\sum_{j=1}^{3} \alpha_j M_j\right)^i + f^X_A (A + K_0)^i + f^X_P (G + P)^i + f^X_T \theta^i T^i \tag{2}$$

where $X$ = HA, AA, or LA. For the high-achieving (HA) students, these coefficients are:

$f_M^{HA} = 0,40$, $f_A^{HA} = 0,30$, $f_P^{HA} = 0,15$ and $f_T^{HA} = 0,15$. For average-achieving (AA) students, $f_M^{AA} =$

0,30, $f_A^{AA}$ = 0,40 , $f_P^{AA}$= 0,15 and $f_T^{AA}$= 0,15 whereas for low-achieving (LA) ones we obtain $f_M^{LA}$= 0,30, $f_A^{LA}$ = 0,20 , $f_P^{LA}$= 0,15 and $f_T^{LA}$= 0,15.

The coefficients $f^X$ can be interpreted as the relative weight that each term of Eq.1 has, and they are chosen so that the average value of the $K_f^i$ calculated for the individuals of each group is as close as possible to the average value of the real final grades of such students. Note that with this interpretation of the coefficients $f^X$, Eq. 2 can be considered as a multiple regression model, as our calculation seeks for minimizing what in the definition of the multiple regression model is called the error variable (which is not explicitly included here).

Note that the adjustment values that best fit the first two groups (HA and AA) are such that $K_f^i$ add up to *10*. A different situation arises in the LA group, as their final knowledge $K_f^i$ cannot be fitted with a combination whose values sum *10*, the maximum possible score. Although there are no real impediments to this happening, we find that the best fit is obtained for values of $f^{LA}$ that result in $K_f^i$=8. This is understood in the light of what is observed in Fig. 1, where the students of this group passed the course with lower scores, and it was an indication to consider that the spectrum of knowledge that they could achieve is less than in the other groups, as they had some difficulties that our model was not taking into account.

**Method 2: Agent-based model of knowledge acquisition**

Here we present a stochastic version of Knowledge Acquisition Model described in Eq. 1. In order to fully understand learning processes, an insight in the temporal unfolding of learning processes in individuals is needed (Steenbeek & Van Geert, 2013). This is the main reason why the agent-based formalism is adequate: the stochastic model is suitable for performing extensive numerical simulations, which allows us to follow the temporal evolution of individual trajectories. The average over several realizations allows us to obtain two types of results: one is the general behavior of the system, based on parameters common to each of the three (HA, AA and LA) groups. The other is the particular behavior of some individuals, using the set of parameters that characterize at each one.

As proposed before, the final knowledge $K_f$ that a student acquires on a given topic is the sum of factors that contributes in the acquisition of it. For this model, we separate the knowledge in four areas: motivation (*M*), aptitude (*A*), influence of the peers (*P*) and influence of the teacher (*T*). We also consider a prior knowledge ($K_0$) for each individual.

We use the fitting values of Eq. 2 for the different groups that we want to simulate and include all the information we collected for the previous one, i.e., the values obtained in the surveys and observations. In this sense, the educational context remains the same.

**General Description**

The agent-based model is composed by *N* agents, whose knowledge evolves in discrete time steps. The time unit is a week, and the duration of a semester corresponds to 16 weeks. Although this temporal discretization is a necessary simplification to perform numerical simulations, it is compatible with the fact that students incorporate knowledge continuously. As in the real case, the maximum value of final knowledge $K^i_f$ that an agent *i* can achieve is *10*. In order to compare the real data with the numerical results, all the simulations presented here are made for *N=81*, from which 13 belong to the HA group, 34 to the AA group and 34 to the LA group.

Each agent is assigned a vector named "knowledge" that has the following components: $\boldsymbol{K}^i(t) = [K^i_0, A^i(t), M^i(t), T^i(t), P^i(t)]$.

For the sake of simplicity, here we consider a single *P* term that includes the *P* and *G* terms from Eq. 2. This means that in our agent-based models we merge both contributions:

the one we receive through the surveys on whether they study alone or in groups (*P*), and the observation about the formation of clusters during classes (*G*).

The component $K_0$ corresponding to prior knowledge is determined randomly within the range of values observed in the actual data, its value is set at the beginning of the simulation and remains constant throughout the process. On the other hand, the other four elements of the vector knowledge are initially null and their value increases with each step of time, if the conditions of incorporation of knowledge are satisfied. The dynamics of the incorporation of new knowledge is related to another vector associated with each agent, that we named "potential". The vector potential has four elements, each one associated to each term of the knowledge, $\boldsymbol{p}^i = (p^i_A, p^i_M, p^i_T, p^i_P)$.

Each element of the vector $\boldsymbol{p}^i$ is a random number between the maximum and minimum values obtained by the students of each group (HA, AA, LA) in each area (*A, M, T* and *P*). These values are chosen at the beginning of the simulation and remain constant throughout the run. The values awarded to each individual determine the maximum knowledge they can obtain in each area at each time step.

Then, at each time step random numbers $r_n$ are sorted for each of the four terms *($P^i$, $T^i$, $A^i$ and $M^i$)* and if they fall below the corresponding potential $(p^i_A, p^i_M, p^i_T$ or $p^i_P)$, a fraction *q* of knowledge from each area is added to the vector knowledge of the agent:

$$A^i(t) = A^i(t-1) + q f_A^X \quad \text{if } r_n < p^i_A, \text{ otherwise } A^i(t) = A^i(t-1)$$
$$M^i(t) = M^i(t-1) + q f_M^X \quad \text{if } r_n < p^i_M, \text{ otherwise } M^i(t) = M^i(t-1)$$
$$T^i(t) = T^i(t-1) + q f_T^X \quad \text{if } r_n < p^i_T, \text{ otherwise } T^i(t) = T^i(t-1)$$
$$P^i(t) = P^i(t-1) + q f_P^X \quad \text{if } r_n < p^i_P, \text{ otherwise } P^i(t) = P^i(t-1) \quad (3)$$

where, again, the subscript *X* (= HA, AA or LA) indicates the group at which the agent *i* belongs, and the $f^X$ are the fitting of the terms obtained with the KA model (see Eq. 2).

From the above it follows that the maximum knowledge that an individual can acquire in a single time step is equal to the sum $[q (f_A^X + f_M^X + f_T^X + f_P^X)]$. This value is the same for all the agents belonging to the same group. However, the chances of being able to reach this maximum value depends on the potential assigned to each individual, and on the value of the random number that changes in each evaluation. In other words, the possibility of adding the contributions of each area at each time step are random events.

The parameter $q$ should be interpreted as a learning rate, since it controls the speed of the process in order to make it compatible with the duration of a semester. More specifically, $q$ is the discretization of the knowledge acquired in each time step, and it was chosen in such a way that, after 16 weeks of evolution, the agents reach the average knowledge of the system we are modeling. Its value is chosen so that the simulations satisfy $<K_f> = \sum_{i=1}^{N} K_f^i$, where $K_f^i$ is the final value of the knowledge obtained for the agent $i$, $K_f^i = A^i(t=t_f) + M^i(t=t_f) + T^i(t=t_f) + P^i(t=t_f)$. We find that the value that meets this condition is $q \approx 0.1$ and depends only weakly on the particular characteristics of the agents.

Due to the complex interactions of individuals within the classroom, dynamics associated with collective behavior emerge in real systems that should be considered in the models. In order to consider the effect of the interaction in a given cluster, the contribution of the pairs to the knowledge expressed in Eq. (3) was modeled in several different ways.

In a first version, that we call ABM0, the interaction between pairs mimics the approach of the KA model: the influence of the pairs is taken in a effective manner, a kind of "mean field" where the contribution of the other students is implicit and there is no information about the agents (nor the number nor their personal characteristics) belonging to the same cluster of the agent *i*. In this approach we only consider that agents have more or less contact (and learning) from their peers, and the level of interaction between pairs is measured by the magnitude of *P*. The ABM0 is useful to compare the results of the KA model, since it is a stochastic version that quite faithfully reproduces the Eq. 2. The implementation of this model is exactly as described in Eq. 3.

One simple way to explicitly take into account the interaction between individuals is to form clusters between the individuals in the course and consider the interaction between them when evaluating the effect of pairs. In the model, this was implemented by randomly selecting at the beginning of each run the partners with whom each agent will interact. In this way, clusters of different sizes and compositions are formed. In this case, the following alternatives were tested:

In the ABM1 version the agents of a given cluster imitate the behavior of the partner who presents the best performance when interacting with the rest of the cluster during the last week. This should be interpreted as that the members of a cluster learn from the one who seems to have more possibilities to incorporate knowledge in the different areas considered. Such a condition is mathematically equivalent to selecting an agent that, at time *t*, maximizes the quantity $[q(f_A^X + f_M^X + f_T^X + f_P^X)]$. At this point it should be remembered that, although the possibility of adding the contributions of each area are random events, agents with higher potential are more likely to be successful in the process because their potential have higher values. Once the agent with best fitness of each cluster is detected, the rest of the agents within the cluster copy the value $f_P^X$ from her/him.

The ABM2 version corresponds to a less restrictive way to learn from the partners in a given cluster. In this case a random agent is chosen and, if it has a value $f_P^X$ greater than its own value, it copies it. This could be interpreted as that the interaction between the individuals of a given cluster is weaker than in the previous case, since they only know the aspect they share and that is reflected in the term of pairs.

Finally, we compare the previous versions with a totally random one (ABM3), in which each agent can copy the $f_P^X$ value of any agent in their cluster chosen at chance. This means that no information about the performance of the partners is actually shared during the interaction.

# RESULTS

## Knowledge Acquisition (KA) model

We present in Fig. 2 a comparison between the final grade for each student and the final knowledge obtained from Eq. 1 (KA model) using the collected data. To ease the comparison, the results are shown in descending numerical value (the order within each group is not related to Fig. 1 but the data are the same). Dotted lines represent the separation between groups of students. The global behavior of the KA model follows the general trend of the data. This behavior is particularly clear for the HA group, as the slope coincides with that of the real data. For the other two groups the trend is not so clear, however the results are compatible. The observed dispersion is due to the presence of particular cases, whose complete evolution is not captured by the model and will be analyzed later.

<<Insert Figure 2 About Here>>

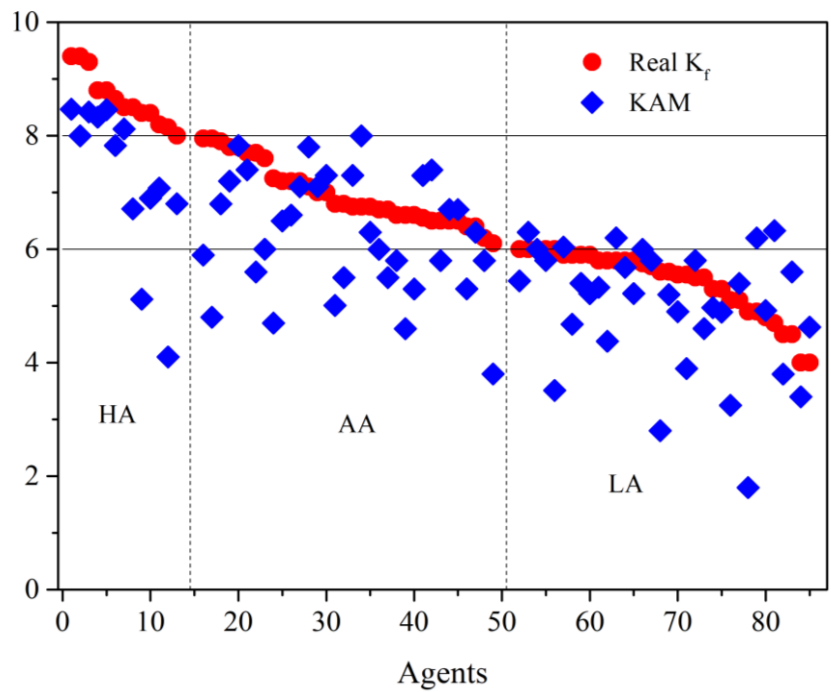

Fig. 2. Final knowledge comparison in descending order between the final grade for each student (red circles) and the collected data using Knowledge Acquisition Model (Eq. 1) (blue diamonds) for *N=81*.

To measure how strong the relationship is between two quantities in our model, we use the correlation coefficient *r*. The calculation of the correlation coefficient was performed by comparing the sum of the items measured in the surveys (the value of $K_f$ obtained with the KA model) with the actual data (the final grade of the students). The value *r=0.7* obtained is high enough to guarantee the correlation between the mentioned quantities.

Although there were variables that were measured several times during the process ($M$, $K_0$, α), the only quantity for which the calculation of Cronbach's alpha (Cronbach, 1951) as a measure of the reliability makes sense was the motivation *M*, the obtained value being 0.63. Given the characteristics of the questions to measure α and $K_0$, we do not expect them to have a correlation due to their lack of similarity; in the case of $K_0$ due to the different topics in the course and in the case of α because they are different questions asked at different times in the process.

It is worth noting that the surveys were carried out with the purpose of obtaining quantifiable information about the variables related to Eq. 1 (aptitude, motivation, previous knowledge, etc.). However, its design does not allow defining scales or measuring the dimension of such variables.

The construction of the KA model is independent of the final grade for each student $K_f^i$ since it comes from the hypothesis that knowledge is the sum of a set of four factors that contributes in its acquisition. The results shown in Fig. 2 indicate that the terms considered in the KA model reproduce approximately the behavior observed for the real data, reflecting the main mechanisms in the process of acquiring knowledge.

## Agent-based models

In what follows we present the results of the agent-based stochastic version of the Knowledge Acquisition model. This complementary model will allow us to partially fill in some of the shortcomings of that model and validate its main results.

**Agent Based Model with effective interaction between pairs**

The simplest version of the agent-based model (AMB0), where the interaction between pairs is introduced as a mean field, allows us to analyze both the collective and individual behavior of agents and compare it with real data.

To study the collective behavior of the individuals, we simulate the behavior of *81* agents distributed in groups with the same proportion and individual characteristics of the real case, i.e., *13* HA, *34* AA and *34* LA students. The results obtained are plotted in Fig. 3, where we represent the average value of final knowledge $<K_f>$ and its dispersion $\sigma_{Kf}$ over $10^3$ independent realizations. This figure shows that the model fits very well to the real data of each group of agents, which seems to be a quite robust behavior for this model and serves as a validation of the chosen methodology.

<<Insert Figure 3 About Here>>

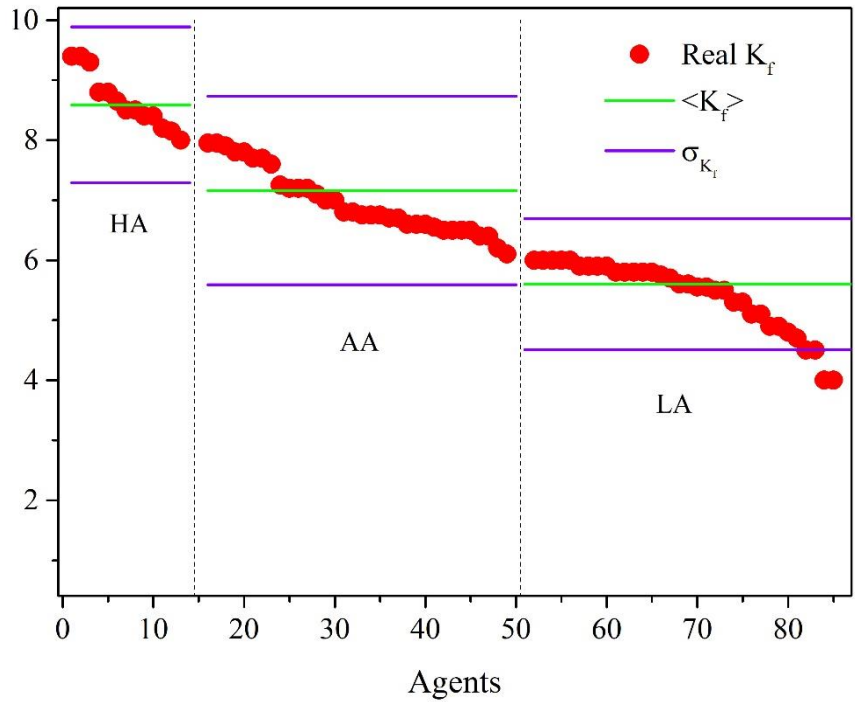

Fig. 3. Agent based model for each group of agents: in solid lines the average value of final knowledge $<K_f>$ (green) and its dispersion $\sigma_{Kf}$ (purple) over $10^3$ independent realizations for each group, and *N=81* (*13* HA, *34* AA, *34* LA). Red circles are the final grade of each student.

This simple model is also useful to study individual cases in the knowledge acquisition process. In the previously described results, we show a global insight about the knowledge acquisition in student groups. However, inter -and intra- individual variability exists, i.e., "differences in the behavior within the same individuals, at different points in time" (Steenbeek & Van Geert, 2013). This variability also tells us about particular situations that exist in any learning process and even more, in the application of surveys and the making of observations in the classroom, an essential part of this work.

Regarding the inclusion of such variability in our model, we perform stochastic numerical simulations with the specific characteristics of each individual, obtained through observations and surveys in the classroom, and compare them with the final grade obtained for the same student. In Fig. 4 we show, for several students from the three groups, a comparison between their final grade and its individual trajectories obtained using the KA model and the ABM0 model. In this figure we present cases in which this comparison between models and data fits very well, and others in which it does not. The comparison was made for the *81* agents studied, although here we only show those that are more representative of what was observed.

<<Insert Figure 4 About Here>>

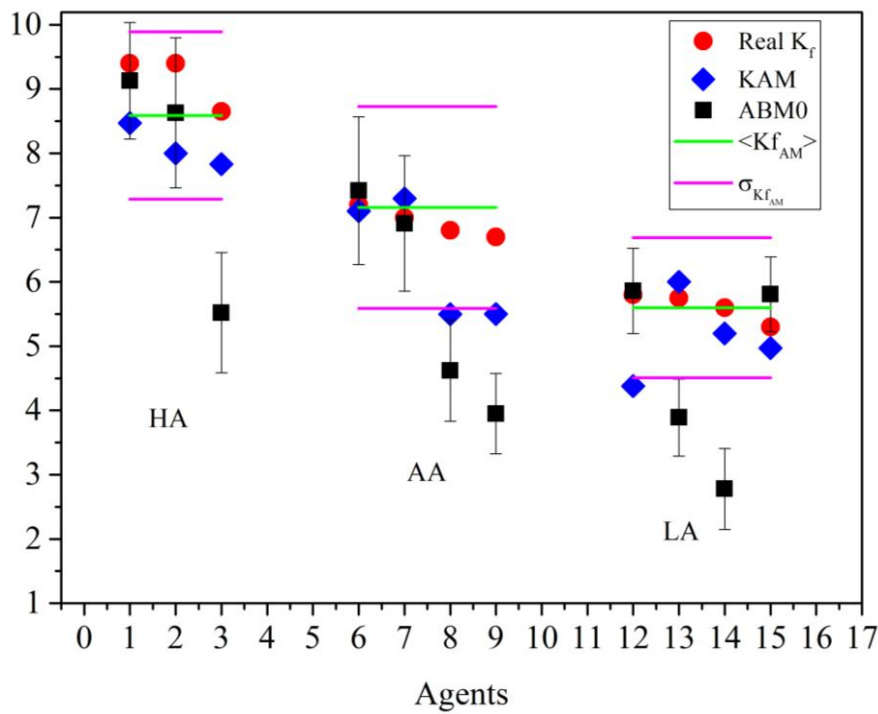

Fig. 4. Comparison of final knowledge for some selected individuals. Final scores (red circles), KA model (blue diamonds) and the ABM0 (black squares) are drawn. The ABM0 model was run $10^3$ times for each agent and its average and dispersion are indicated with solid lines.

The cases in which the agent-based model does not reflect the result obtained by the corresponding student were analyzed individually, following its evolution throughout the semester of classes. Interestingly, in all the cases, regardless of which group they belonged to, the surveys of these students showed low prior knowledge and very low levels of motivation. Furthermore, in such cases the ABM0 model underestimated the final result, indicating that some successful strategies used by students are not contemplated in this model and reveals the variability in knowledge acquisition of some agents.

**Agent Based Model with clusters**

One way to render the model more complex and realistic is to explicitly include the effect of the interaction of pairs that belong to the same cluster. The formation of clusters is related to the observations of the spatial distribution of the students during the course of the class. In general, the students maintained this distribution during the semester, from which it can be deduced that there are groups that interacted more closely than with the rest of their peers. The simulations that we present in Fig. 5 correspond to the version ABM1, that incorporates the structure of clusters. In this case, the individuals of a given cluster copy the value of the pairs term *P* of the agent that was able to maximize their learning in the last time step. The possibility of carrying out many realizations for each set of parameters makes it possible to estimate which is the most probable scenario in each situation analyzed. We made $10^3$ realizations for a fixed number of clusters indicated in the horizontal axis with different seeds. Each cluster is composed of randomly chosen individuals and the fraction of knowledge added at each time step, which is the only free parameter used to fit simulated and real data, is *q = 0.105*.

<<Insert Figure 5 About Here>>

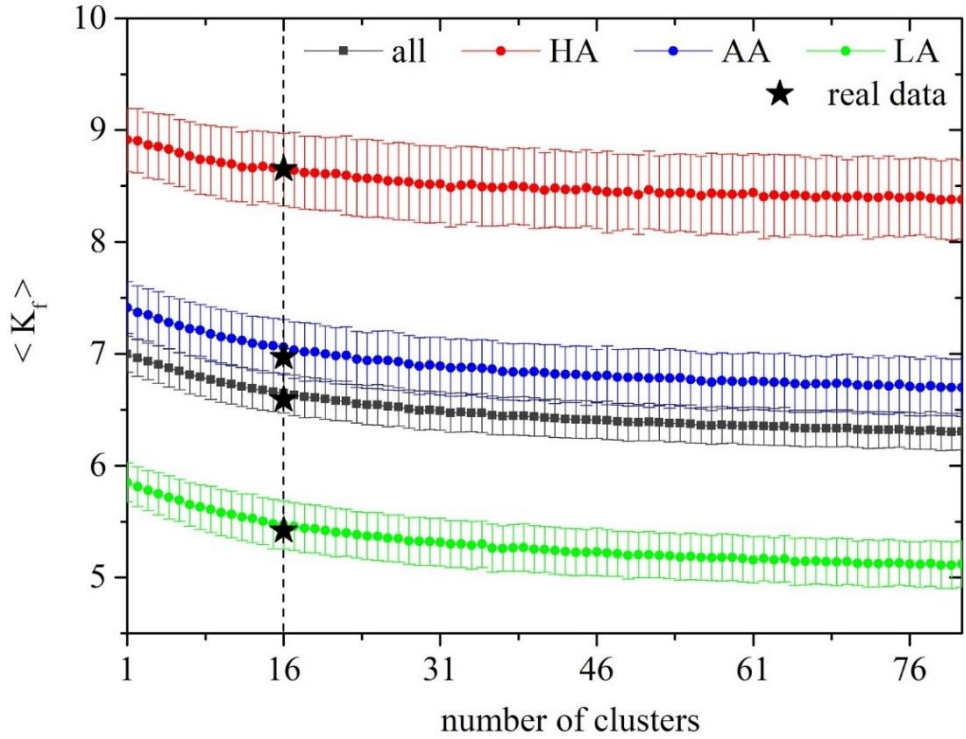

Fig 5: Simulations with the ABM1 model for *N=81* (*13* HA, *34* AA and *34* LA agents). For a fixed number of clusters formed at random, indicated in the horizontal axis, $10^3$ realizations with different seeds were made. The average knowledge $<K_f>$ of the all agents together is indicated in gray squares and the average knowledge of each group (HA, AA and LA) is plotted with red, blue and green circles, respectively. Error bars indicate the standard deviation of each point. Black stars indicate the real data obtained averaging the final grade of the *81* students, which are distributed in *16* clusters.

Several things should be noted in Fig. 5. One is that the real case is very well reproduced for the model, as the average final grade of the students fall within those obtained with the simulations for a number of clusters coinciding with that of the real data. Another significant result is that final knowledge is always greater when individuals form clusters (the bigger the better) than when they work individually (what corresponds to having a number of clusters equal to *N*). The difference between the two extremes of the gray curve of Fig 5, which measures the average final knowledge of the 81 agents together, is 6.9%. This is a way to know how much students benefit from belonging to a large cluster compared to studying alone. Another characteristic observed, that allows quantifying the behavior of the HA, AA and LA students, is the difference in the knowledge acquired for each group separately at the two limiting cases just mentioned. We found that the LA students are most favored by the interaction, with an improvement of about 7.3% when forming a large cluster. A similar improvement of ~7.2% was also found for the AA group. But more importantly, HA students also have an improvement, although in lower percentages (~5.3%). These numbers change when courses of different sizes and proportions of HA, AA and LA individuals are simulated, but the relative behavior between groups remains the same. The rest of the points shown in Fig. 5 helps to predict what the average final result of the students would be if they were distributed in clusters of larger or smaller sizes. An analysis regarding the advantages of collaborative learning will be done in the Discussion Section.

It is important to note that the new results with ABM1, which explicitly considers the formation of clusters, do not detract from the mean-field version (ABM0), as shown in Fig. 6. Although the advantage of ABM1 is the possibility of predicting the behavior of the system for different scenarios, the results of both models are very similar when it comes to reproducing the $K_f$ ranges of the real data.

<<Insert Figure 6 About Here>>

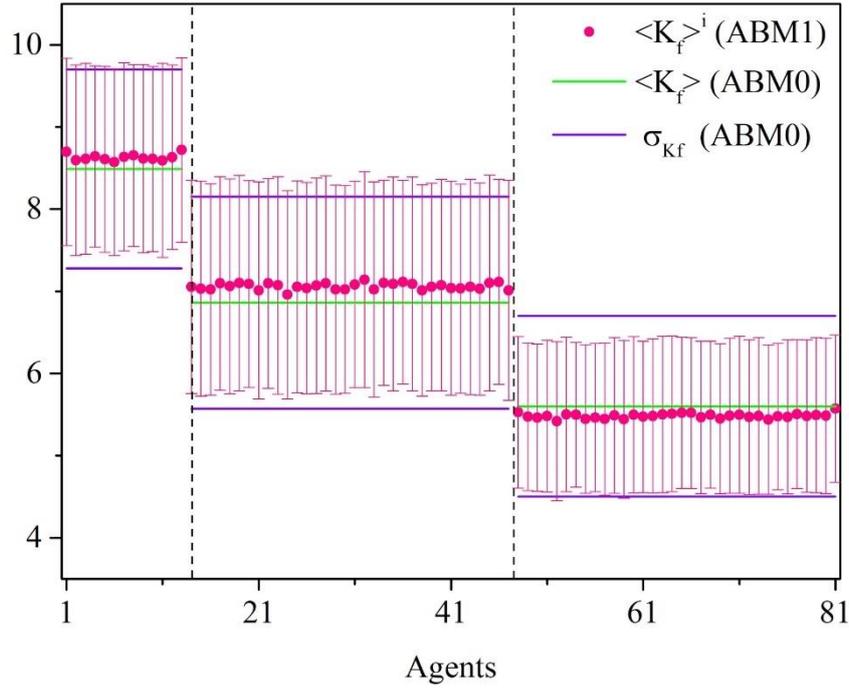

Fig 6: Comparison of ABM0 and ABM1. Dots: average final knowledge $<K_f>$ for the ABM1 for *N*=81 agents distributed in 16 clusters formed at random. Error bars indicate the standard deviation of each point. Solid lines: average value of final knowledge $<K_f>$ (green) and its dispersion $\sigma_{Kf}$ (purple) for the ABM0. In all the cases, $10^3$ independent realizations were made for each agent.

The results presented in Fig. 5 on the improvement of average knowledge when working in clusters may be thought to originate from the fact that the constitution of the clusters was done at random, since in any cluster large enough there will be HA individuals that can be copied by the less favored. To check if this is the main cause of what was observed, simulations were carried out with "ordered" clusters, in the sense that are mostly integrated with individuals of the same group, HA, AA or LA. These simulations gave qualitatively similar results (not shown here), with curves presenting the same tendency as those of Fig. 5. The only notable difference is that the slope of the curves is greater, which is reflected in the fact that the difference between the two extremes of the curves is greater than in the case of random clusters: 9.1% for the entire group, 6.9% for HA, 9.4% for AA and 9.7% for LA students.

Another possible explanation of the previous results could be that there will always be an advantage in copying the behavior of the most successful individual in the cluster. This could be evaluated by comparing what happens with the ABM2 model, where agents copy the value of an agent who only did better that day in the term of pairs (and not in the set of all possible learnings). Again, ABM2 has the same qualitative behavior as ABM1, both when the clusters have random compositions and when they are organized in order, according to groups (not shown here). The improvement in knowledge by studying in a single cluster compared to doing it individually is 6.9% for the whole group, 5.7% (HA), 7% (AA) and 7.2% (LA) if the clusters are formed at random. When assembled in ordered clusters according to their performance, these values are respectively 10.3%, 8.4%, 10.5% and 10.7%.

A different behavior is obtained for the fully random case (ABM3), which presents a flat curve without significant differences for different cluster sizes, both when they are assembled randomly and when they are assembled in an orderly manner. This means that working in a group improves learning as long as some overcoming characteristic of the group is taken consciously. The fact that the curve is flat for ABM3, which means that there are no advantages to studying in a group compared to doing it alone if there is no collective learning, will be made clear in our analysis below. For completeness, we report the differences between knowledge for a single cluster and individual work. For random clusters formation, 0.09% (all), -0.76% (HA), 0.09% (AA) and 0.41% (LA). For ordered clusters, 0.06% (all), -1.23% (HA), 0.10% (AA) and 0.51% (LA).

Beyond the fact that the trend of the curves for the ABM1 and ABM2 and for both, ordered or random clusters, is always as observed in the curves of Fig. 5, it is important to investigate if there are significant differences in these curves. We plot in Fig. 7 the difference between the two extremes of each curve for all the cases previously commented: the three ABM versions of the model with clusters, and for each of them, the ordered or random distribution of the agents in the clusters.

Note first that ABM1 and ABM2 present significant differences between the knowledge acquired by the entire group (Fig 7a) when they work collaboratively (forming a single cluster) than when they do it individually. This follows from the non-overlapping of the averages and their corresponding dispersion bars. On the contrary, the ABM3 model does not present significant differences for either of its two versions of cluster construction (random or ordered).When the behavior of each group is analyzed separately (Figs. 7b, 7c and 7d), a partial overlap is observed in the HA group in the cases in which the clusters were formed randomly but not for the ordered clusters. Groups AA and LA show significant differences in all cases, both for the ABM1 and ABM2 models.

Note that what we commented when describing Fig. 5 is verified: the improvement of average knowledge for the students of the LA group (Fig. 7d) is greater than that of the HA (Fig. 7b). This can be measured as the difference between the extreme values of each case, which are precisely the data shown in full and empty circles.

<<Insert Figure 7 About Here>>

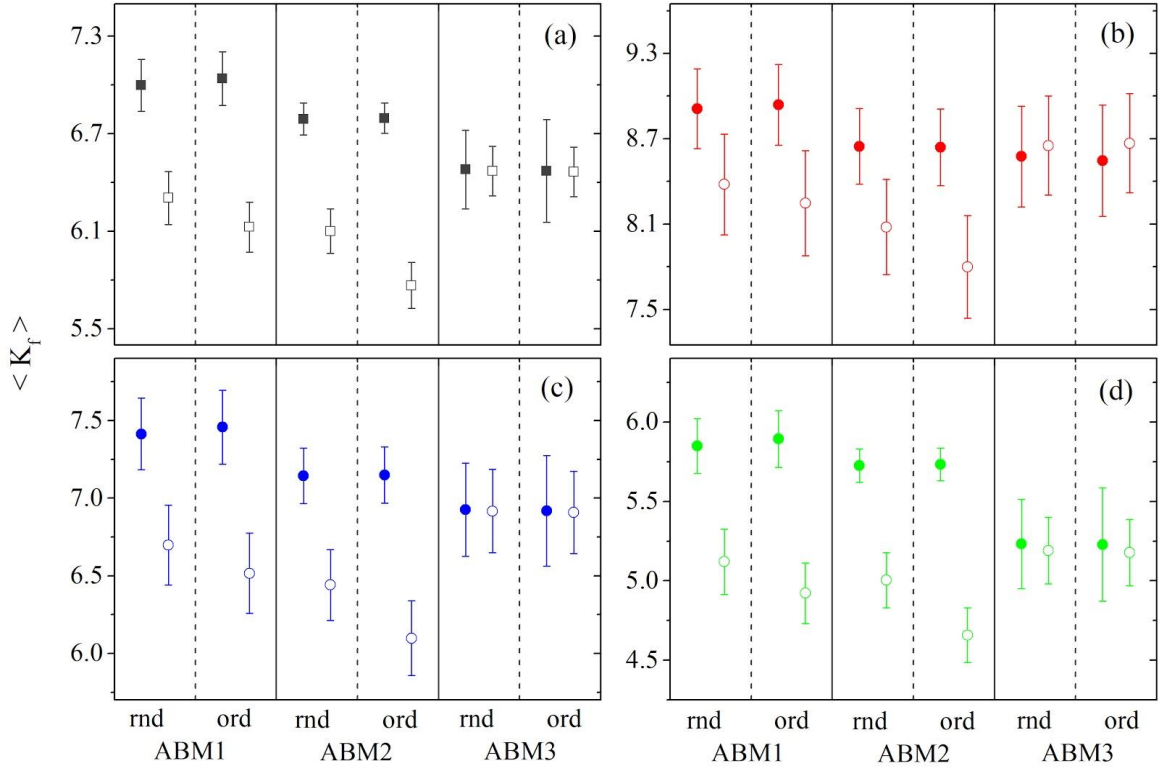

Fig. 7: Average knowledge $<K_f>$ in two extremes cases: one single cluster (full symbols), and individual behavior (open symbols). (a) $<K_f>$ of all agents together, (b) $<K_f>$ of the HA agents, (c) $<K_f>$ of the AA agents, (d) $<K_f>$ of the LA agents. The letters "rnd"(random) and "ord" (ordered) indicate the way the clusters were formed in relation to the HA, AA and LA groups. Significant differences are observed in almost all cases for the ABM1 and ABM2 models.

## DISCUSSION

We presented an analysis of the knowledge acquisition process in a teaching-learning scenario that take place within the classroom using two different and complementary approaches: classroom observations combined with surveys, that we used to construct what we call the Knowledge Acquisition model, and a series of stochastic Agent-Based models, a standard tool of Statistical Physics widely used to study complex systems. The integration of these diverse tools is the aim of this work, in an attempt to analyze the process of acquiring knowledge from a multidisciplinary perspective.

As a first step, we designed surveys whose objective was to obtain information about the mechanisms that we proposed as relevant during the knowledge acquisition process, and that we divided into areas. Thus, the surveys carried out three times during the semester of classes inquired about motivation, skills and prior knowledge, their relationship with teachers and the strategies used to interact with them, and finally, interaction with peers at the time of studying out of the classroom. This last item was complemented with observations within the classroom, where the spatial distribution of the students was recorded and thus information on the size and distribution of the students in what we call "clusters" was collected. On the other hand, at the end of the semester the final grade obtained in the course was recorded for each student. Given that the subject taken corresponds to the area of Physics, the final grade obtained is a good proxy for the performance of the students. Although in this approach we associate learning with knowledge, which is measured through grades, we believe that this proposal is valid because we are evaluating a specific type of learning, whose content is scientific. With this we do not suggest that this is the only valuable thing that is learned in the classroom, but that we intend to collaborate with the understanding of a complex process through the tools that we handle.

Following a previous work of Bordogna and Albano (2001) we classified the students into three groups according to their performance (HA, AA and LA), see Fig. 1. This was very useful to analyze the construction of knowledge of the students belonging to each of the three groups, since we found common characteristics in each of them.

The information collected in the surveys and observations were integrated into the first method described in this work, which we call Knowledge Acquisition model (KA model). The adjustment of this model with the final achievements of the students allowed us to associate the different groups of students into groups with very different characteristics and behaviors. For each group we find a fit of the terms we use to describe knowledge acquisition at both individual and global level. This tells us that the choice of the terms that we consider to be relevant when analyzing the different contributions to the acquisition of knowledge was correct. Of course, this simplification does not take into account a myriad of variables that are integrated to give rise to the unique process that each person experiences, but it allows us to validate our choice as one that rescues the main contributions common to all individuals. In Fig. 2 we show that the general behavior of individuals can be described with the KA model equation, which is simply the sum of the relative contributions of each of the proposed mechanisms. This result also indicates that the information collected in the surveys and observations was sufficient to construct an adequate representation of the process.

    The second method used in this work belongs to the family of Agent-Based models. We design versions with different degrees of interaction between individuals to model various possible scenarios. A first version, which we call ABM0, corresponds to considering the interaction between pairs as an average value (or a "mean field") that is different for each agent but that does not take into account the structure of the cluster to which that individual belongs. With this simple model we carried out a large number of simulations with the same conditions of the real system in terms of the number of individuals studied and their distribution in three groups according to their achievements (HA, AA and LA) and we found that the average values and dispersions were in very good agreement with the real data (Fig. 3). We were also able to model particular individuals, assigning to the parameters the values obtained from the surveys and observations of that student. The comparison between the real data and the simulations leads to an interesting analysis, as was presented in Fig. 4. We observe some cases where the coincidence between both is very good and others in which the ABM0 does not reflect the result obtained by the corresponding student. In the latter cases, regardless of the group to which they belong, we find that their surveys indicate low prior

knowledge and very low levels of motivation. According to Greeno et al (1996) all of the psychological perspectives on learning school subjects assert that learning requires the active participation of students and for this the motivation from any of its perspectives is necessary. This is clearly reflected by the model. Besides, in all these cases the ABM0 underestimated the final result obtained by the students. We believe this would indicate that there are successful strategies used by students that were not considered in this model and should be explored. The results presented in Fig. 4 helps to understand the complexity of process behind an exam score and reveals the variability in knowledge acquisition of some agents. This opens, among others, possibilities to improve teaching and communication strategies with students, in order to have an early diagnosis of their needs and thus improve the process of acquiring knowledge.

      With the aim of deepening the study of collaborative learning and its effects on learning trajectories, we developed more complex versions of agent-based models. We focus on analyzing the effect that the structure of study groups (here called clusters) has on the construction of knowledge. We proposed three different options that model different peer learning mechanisms. The ABM1 corresponds to the case with the highest interaction, since the individuals in a given cluster check which of their peers is the one that currently has the best performance in all the areas considered, and they copy the part corresponding to the interaction between peers. Here it is important to note that taking a complete copy of the knowledge of the most successful individual it is not a good proposal, since it is unrealistic to think that the aptitudes or motivations can be simply "copied", and furthermore, would lead to an unrealistic situation in which all agents end up having the same final knowledge. This model was simulated in two different versions in relation to the way the clusters were assembled. The random version corresponds to forming clusters with random individuals, regardless of which group (HA, AA, LA) they belong to. In this case we find that when the

whole group works collaboratively, the average final knowledge of the students is significantly higher than when they do it individually (see Fig. 5). When the calculation is done separately for the three groups, we found that, although all the agents benefit from the interaction, those with the lowest performance benefit the most. However, in this collaboration everyone wins, since those with the best performance (HA) are also favored. The second version of ABM1 corresponds to the distribution of individuals in the clusters in an orderly manner. This means that they are formed according to their performance, and so the clusters are mostly composed of individuals from the same group. In this case it is observed that the improvements are even greater for all groups. This counter-intuitive result, in which the lowest performing individuals individuals (LA) are favored even when the cluster to which they belong is made up of individuals of similar performance, can only be explained in the context of a complex system in which interactions between individuals result in something more than the linear sum of individual behaviors.

The results presented in Fig. 5 are very important since they not only to validate the model based on the good agreement obtained between the real data and those of the numerical simulations, but also allows us to predict what the performance of the students would be if they were integrated into clusters of different sizes and distribution of individuals according to performance. This result reinforces the importance of collaborative work (Webb, N. M. & Palincsar, A. S, 1996). Moreover, Fig. 6 shows that the results obtained with the more complex model, ABM1, which explicitly considers the formation of clusters, do not go against the mean-field version, ABM0. The advantage of ABM1 is, again, the possibility of predicting the behavior of the system for different scenarios.

The qualitative behavior of the curves in Fig. 5 is repeated for the version of the Agent-Based models that we call ABM2. This version corresponds to a situation in which the interaction between individuals of the same cluster is weaker. In this case, the agents only

*Knowledge acquisition as a complex system* 36know the performance of their partners in terms of their pairs' learning, and they randomly copy one of them who performs better than their own. For ABM2 we also consider two versions in the conformation of the clusters (random or ordered). The improvement in knowledge by studying collaboratively compared to doing it individually is similar to the ABM1, both if the clusters are formed at random or when assembled according to the agent performance. Again, it is always better to work in collaborative clusters if the opportunity is used to learn from each other.

      Finally, the previous results were contrasted with the ABM3 version of the model, where no learning is considered but only a copy at random regardless of the performance of the interacting agents. In this case there are no advantages in the formation or not of clusters. In Fig. 7 we resume the main results of the three agent-based models with clusters (ABM1, 2 and 3) and show that in almost all the cases we observe significant differences between the two extreme cases simulated: the conformation of a single big cluster with all the students interacting and learning from each other, and the individual learning.

      Although the exact mechanism linking students' experiences in classroom groups to their learning, conceptual development, and social-emotional outcomes are complex and not yet well understood, few would dispute that group interaction and mediating processes have major influences on outcomes of group work as mentioned by Webb et al (1996).

      The results obtained in this work reinforce, among others, the ideas studied through the theory of sociocultural learning within the classroom, highlighting the importance of the social environment of individuals as well as mutual collaboration for the acquisition of knowledge. All this was done using novel methods that were not associated with educational issues even knowing the complexity of the interactions that occur there.

      We believe that this work proposes a complementary way of understanding the dynamics of the classroom and its impact on learning compared to conventional empirical

approaches. Along with observing and collecting information on the determining factors in the knowledge acquisition process, this approach adds less tangible elements, originating from the underlying complexity of a system composed of interacting individuals. The formalism on which these models are based allows us to equate the learning process to that of a system of interacting individuals that gives rise to non-linear dynamics. Thus, it is possible not only to reproduce real situations observed, such as the improvement of academic performance when collaborative learning occurs, but also to quantify the degree of improvement that can be achieved. Finally, this approach allows proposing different work formats in the classroom, according to the number and characteristics of the students, in order to promote the exchange of knowledge and experiences and improve the overall performance of the group.

As a future work, besides adding more and better questions to the surveys to complete the proposed Knowledge Acquisition Model, it might be worth studying the behavior of the models considering actors in another educative contexts with the aim of propose improvements to the methods and skills that lead to a better acquisition of knowledge.

## ACKNOWLEDGMENTS

F. Velásquez-Rojas and M.F. Laguna acknowledge support from Daniela Zacharias for their advice on the calculation and handling of reliability scales as well as the interpretation of the results of the surveys carried out. We also want to thank Marcelo Trivi, Associate Professor of the course of Physics II at the FI - UNLP for providing the specific educational context during the performance of this work.